# SELF-GENERATED OFF-AXIS HOLOGRAPHY IN INTERFEROMETRIC OUT-OF-FOCUS IMAGING OF ICE PARTICLES

MOHAMED TALBI,[1] GÉRARD GRÉHAN,[1] MARC BRUNEL[1,*]

[1]UMR CNRS 6614 CORIA, Avenue de l'Université, BP 12, 76801 Saint-Etienne du Rouvray
*Corresponding author: marc.brunel@coria.fr

**Interferometric out-of-focus imaging of ice particles is realized. The overlapping defocused images of two nearby ice particles are analyzed. If one particle is much smaller than the other one, the pattern constitutes an off-axis hologram whose analysis gives the exact size and shape of the biggest particle. Using two angles of view, we obtain a 3D description of the biggest ice particle.**

http://dx.doi.org/10.1364/OL.99.099999

Interferometric out-of-focus imaging enables the characterization of spherical droplets or bubbles in a flow with applications in sprays, combustion, meteorology, fluid mechanics [1-8]. The technique can be extended to the characterization of irregular rough particles. Interferometric out-of-focus images of such particles are then speckle-like patterns. Their analysis enables the determination of quantitative informations about the morphology of the particle [9-14]. Based on this property, some information concerning the 3D-shape, and orientation of the particle can be obtained using a multi-view set-up [15]. An important application is aircraft safety. Detection and size measurements of ice crystals in the atmosphere are indeed crucial. Recent works showed that the technique can be used in the case of ice particles, with a size measurement error rate for single particles reasonable [16,17]. Nevertheless, the analysis of overlapping images necessitates the development of specific methods for higher concentrations in particles [18,19].

There is no theoretical model that can predict rigorously the interferometric out-of-focus images of rough particles of any shape and texture. Fortunately, recent experiments have validated a simplified approach: assimilating an irregular rough particle illuminated by a laser to an ensemble of Dirac emitters located on the envelope of the particle, a scalar expression of the electric field received by the camera can be expressed [9,10]. It is then shown that the 2 dimensional Fourier transform of the interferometric pattern gives the contour of the 2 dimensional autocorrelation of the initial repartition of the emitters, i.e. of the initial illuminated particle [12]. All experiments realized confirmed this result [15-19]. This was further corroborated by experiments using "programmable rough objects" created on digital micromirror device [20].

When the out-of-focus images of two particles overlap, the initial object is a two-components object. If separation between them is higher than their size, the 2D-autocorrelation of this 2-components object is composed of three large spots: a central one corresponding to the superposition of the 2D-autocorrelations of both particles considered separately, a spot corresponding to the cross-correlation between both particles and its symmetric spot. Finally, if one particle is much smaller than the other one, the cross-correlations reduce to the exact shape of the biggest particle. This case is predicted theoretically, but its observation is very difficult [19]. Intensity scattered by a droplet is proportional to $d^2$ where d represents its diameter. If one particle is 20 times bigger than the other one, the ratio between intensities scattered by both particles should thus be around 400. The observation of interference fringes between both signals and their analysis is then difficult. However, in the case of ice crystals, a pure reflection on a facet of a small particle can significantly reduce this ratio. In this article, we realize corresponding experiments and we show that it is possible to obtain the exact shape of a particle in this configuration.

The experimental set-up is presented in Fig. 1. Liquid water droplets of random size are injected at the top of a freezing column [16]. The temperature inside the column is around −45°C. Droplets freeze during free fall and become ice particles. 4ns, 5mJ, 532nm pulses are emitted by a frequency-doubled Nd:YAG laser. They are sent onto the freezing column through a BK7 window. A set of two out-of-focus imaging lines enables interferometric imaging of irregular particles from two perpendicular angles of view. In addition, two in-focus imaging lines are mounted for both same angles of view. This is obtained with two beam splitters. In-focus images are obtained by using far field objectives provided by ISCOOPTIC (fields of view 2.45mm x2.45 mm, depth of field 1mm). Out-of-focus systems consist of Nikon objectives (focal length of 180mm) with an extension tube providing out-of-focus imaging. The set of sensors are first adjusted to obtain an in-focus image of the same point of the laser sheet in the column. Then, the out-of-focus optical systems (camera + extension tube + lens) are translated forward by Δp=15 mm (out-of-focus imaging). In summary, four CCD sensors are synchronized on the laser pulse for synchronized acquisitions. For both angles of view, the in-focus and interferometric out-of-focus images are recorded. On the top view of fig. 1, cameras 1 and 2 are on an imaging axis that makes an angle of θ=45° on the left while cameras 3 and 4 are on an imaging axis that makes an angle of θ=45° on the right. The dimensions of the CCD sensors used for the in-focus images are 2048 x 2048 pixels (pixel size: 5.5 µm). The dimensions of the CCD sensors used for the out-of-focus images are 1920 x 1200 pixels (pixel size: 5.86 µm). As multi-views experiments will be reported, The reference frame (x,y,z) is presented on Fig. 1 for clarity. After 2D-Fourier transforms of the interferometric images, the corresponding coordinates in spectral domain will be noted (u,v,w).

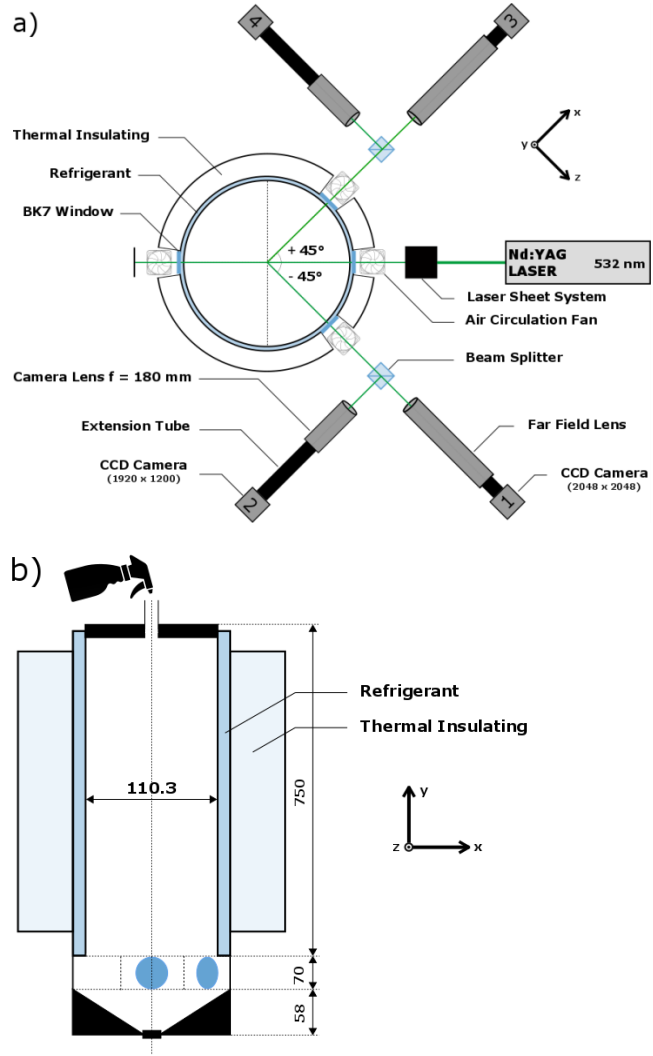

Fig. 1. a) Experimental set-up, top view, b) side view of the column

Let us now present experimental results. Fig. 2(a) shows the in-focus image recorded by sensor 1 while Fig. 2(b) shows the in-focus image recorded by sensor 3 simultaneously. Both figures present a pair of two ice particles. The case observed is very specific: one particle is much bigger than the other one, as presented in the introduction. Fig. 3(a) shows now the interferometric out-of-focus image recorded simultaneously by sensor 2 (same angle of view as sensor 1), while fig. 3(b) shows the out-of-focus image recorded simultaneously by sensor 4 (same angle of view as sensor 3). For quantitative analysis, Fig. 4(a) shows the 2-dimensional Fourier transform of pattern presented in fig. 3(a) and fig. 4(b) presents the 2-dimensional Fourier transform of pattern presented in fig. 3(b). In these two last plots, the scaling factor of both horizontal and vertical axes is coefficient $\lambda B_{tot}$, where $\lambda$ is the laser wavelength, and $B_{tot}$ the B-transfer matrix coefficient of the total transfer matrix that describes propagation from the particle to the CCD sensor for the out-of-focus imaging lines. Note that $B_{tot}$ is not exactly the same for simultaneous observations of sensors 2 and 4. Due to the set-up (angle of view of ± 45° with a laser sheet), $B_{tot}$ has indeed to be determined precisely for each particle, as mentioned in [15]. In the present experiments, $B_{tot}$ is in the range [0.02196 m ; 0.02398 m], depending on the 3D-position of the particle in the fields of view. For both angles of view, we observe three spread spots: the central one (red square) reproduces the superposition of the 2D-autocorrelations of both particles separately. The two symmetric spots (one of them is in the dashed green square), reproduce the cross-correlation of the biggest particle by the smallest particle. As one particle is much smaller than the other one, this spot gives then directly the exact shape of the bright part of the biggest ice particle. Noise and loss of contrast are introduced by the combination of interferometric imaging, 2D-Fourier transforms operation, and by the fact that the smallest particle is not perfectly punctual. Nevertheless, the shapes obtained for both angles of view are quantitatively in relatively good agreement with simultaneous in-focus images already presented in Fig. 2(a) and 2(b). When one particle is much smaller than the other one, it can be viewed as a punctual emitter that acts as a reference wave to generate the off-axis hologram of the second particle [21]. The exact size and shape of the biggest particle can then be directly obtained from the analysis of the interferometric out-of-focus image, as the reconstruction of an off-axis hologram.

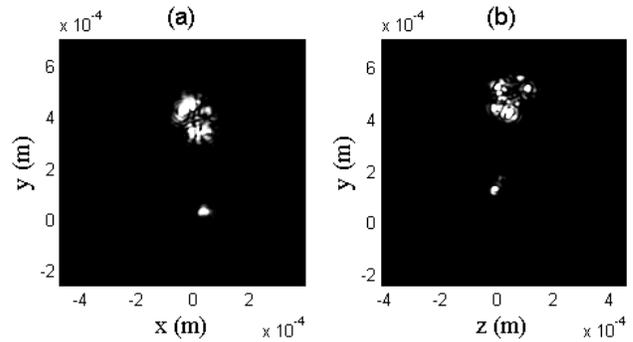

Fig. 2. In-focus images of a pair of ice particles observed with sensor 1 (a) and sensor 3 (b).

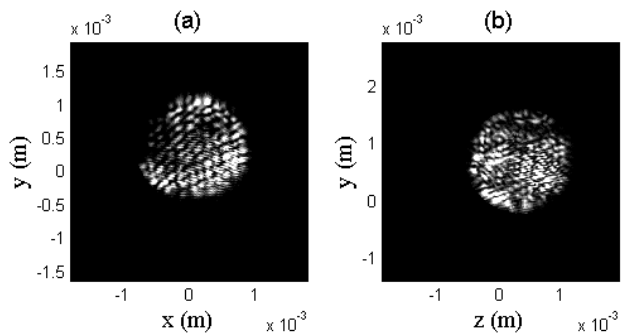

Fig. 3. Out-of-focus images of a pair of ice particles observed with sensor 2 (a) and sensor 4 (b).

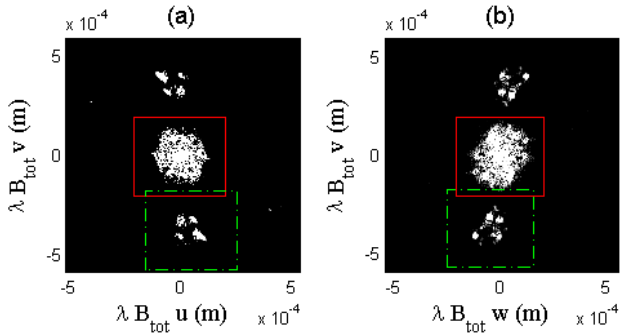

Fig. 4. 2D-Fourier transforms of the patterns observed in fig. 3(a) (a) and in fig. 3(b) (b).

Let us imagine that we observed only the out-of-focus images, (without in-focus imaging lines). Would it be possible to identify that one particle is much bigger than the other one and that we obtain its exact shape ? Fortunately, one criterion can be found: in this case: three spread spots must be observed if separation between both particles is sufficient. And the central spot must correspond with the 2D-autocorrelation of one of these symmetric spots. The superposition of the 2D-autocorrelations of both particles considered separately reduces indeed to the 2D-autocorrelation of the biggest one if the second particle is very small. To illustrate this, Fig. 5(a) shows the central spot of Fig. 4(a) (in red square in Fig. 4(a)), while Fig. 5(c) shows the 2D-autocorrelation of one of the symmetric spots of Fig 4(a) (in the dashed green square in Fig. 4(a)). The same operation is done on Fig. 5(b) and 5(d) from Fig. 4(b) for the second angle of view. We can note that Fig 5(a) and 5(c) match quantitatively well in size and shape. The same comment can be done from Fig 5(b) and 5(d), considering the second angle of view of the particles.

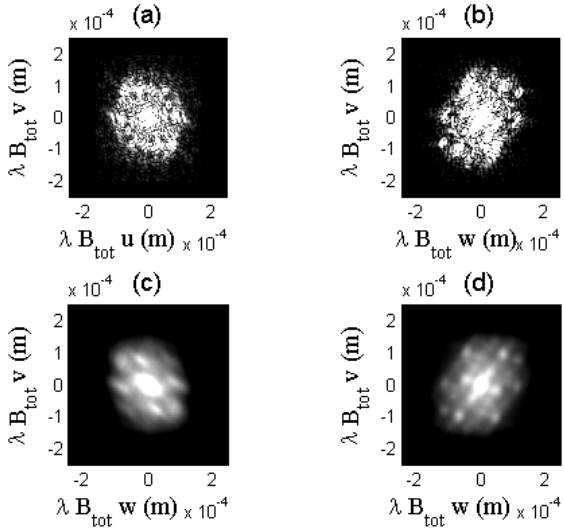

Fig. 5. Central spot of Fig. 4(a) (a) and of Fig. 4(b) (b); 2D-autocorrelations of spot in dashed square of Fig. 4(a) (c) and of Fig. 4(b) (d).

Let us now consider other examples, and discuss sources of noise encountered. Fig. 6(a) shows the in-focus image recorded by sensor 1 while Fig. 6(b) shows the in-focus image recorded by sensor 3 simultaneously. Both figures present a pair of ice particles. Interferometric out-of-focus images recorded simultaneously by sensors 2 and 4 are not reported. Fig. 7(a) and 7(b) show directly the 2-dimensional Fourier transforms of these patterns for both angles of view. The correspondence between the spread spot in the dashed green square of Fig. 7(b) with the corresponding in-focus image of the biggest particle (Fig. 6(b)) is quantitatively very good in size and shape. Nevertheless, a similar comparison between Fig. 6(a) and 7(a) is not so good. This case illustrates the difficulty of a correct reconstruction when one extremity of the illuminated particle appears much brighter than the other extremity as observed on Fig. 6(a) (the sensor is strongly saturated in this case). The contrast of the darkest part is small. It had been already mentioned in reference [15] that the existence of two angles of view enables to identify if roughness and brightness of the particle is such that particle measurement will be correct. One criterion is indeed that the sizes deduced from the two views is the same along the axis that is common to both angles of view (axis y in the present experiment, see Fig. 1).

Let us consider a last example. Fig. 8(a) shows the in-focus image recorded by sensor 1 while Fig. 8(b) shows the in-focus image recorded by sensor 3 simultaneously. Both figures present again a pair of ice particles. Interferometric out-of-focus images recorded simultaneously by sensors 2 and 4 are not reported. Fig. 9(a) and 9(b) show directly the 2-dimensional Fourier transforms of these patterns for both angles of view. The correspondence between the spread spot in the dashed green square of Fig. 9(b) with the corresponding in-focus image of the biggest particle (Fig. 8(b)) is quantitatively very good in size and shape. Nevertheless, a similar comparison between Fig. 8(a) and 9(a) is not possible. The reason is different in this case: the two particles are not sufficiently separated on this view. There is overlapping between each symmetric spot and the central spread spot of the 2D-Fourier transform of the interferometric pattern. The sole Fig. 9(b) would have been interpreted as a unique particle. This last example shows the importance of a multi-view set-up to avoid an erroneous interpretation.

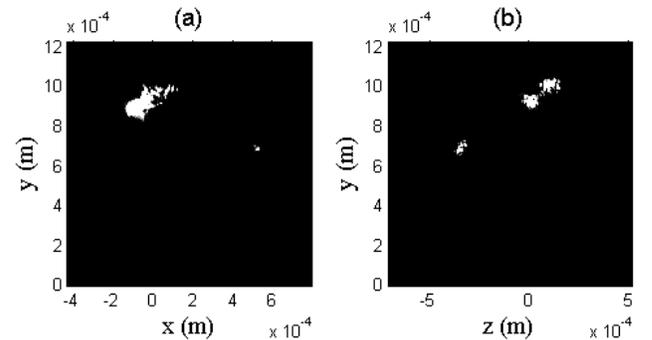

Fig. 6. In-focus images of a pair of ice particles observed with sensor 1 (a) and sensor 3 (b).

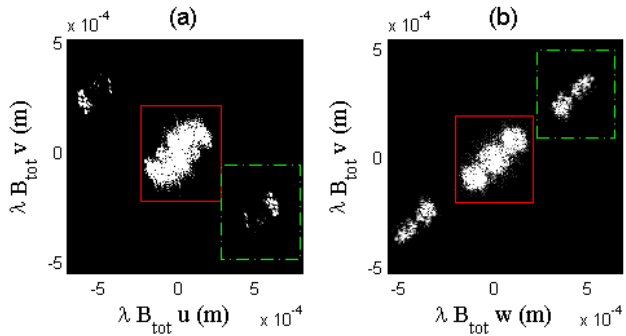

Fig. 7. 2D-Fourier transforms of the interferometric out-of-focus patterns corresponding to Fig. 6(a) (a) and in fig. 6(b) (b).

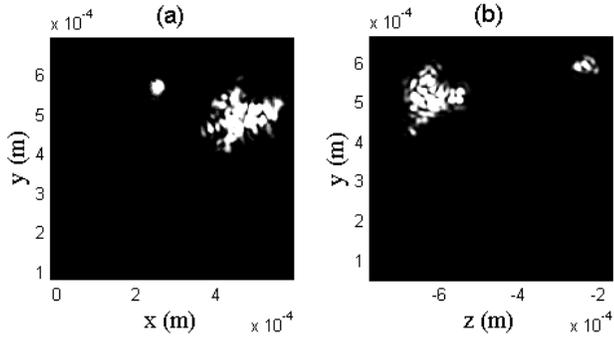

Fig. 8. In-focus images of a pair of ice particles observed with sensor 1 (a) and sensor 3 (b).

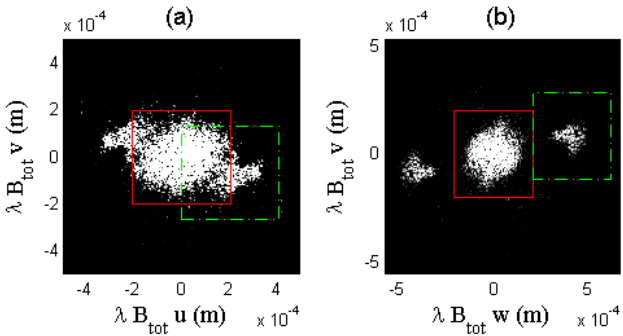

Fig. 9. 2D-Fourier transforms of the interferometric out-of-focus patterns corresponding to Fig. 6(a) (a) and in fig. 6(b) (b).

An ensemble of conclusions can be drawn. These results give another confirmation that the 2 dimensional Fourier transform of the interferometric defocused image of a rough particle gives the contour of its 2 dimensional in-focus shape. The interferometric out-of-focus images of two particles can overlap. When one particle is much smaller than the other one, it can be viewed as a punctual emitter that acts as a reference wave to generate the off-axis hologram of the second particle. The exact size and shape of the biggest particle can then be directly obtained from the analysis of the interferometric out-of-focus image. This case that had been theoretically predicted is observable with ice particles as direct reflections can occur to produce an intense signal on the sensor, although one particle is much smaller. A necessary (but not-sufficient) criterion exists to identify this case: after 2D-Fourier transform of the interferometric image, three spread spots are observed and the central spot must correspond with the 2D-autocorrelation of one of these symmetric spots. The presence of two angles of view enables to avoid errors in interpretation: if one view indicates that there is only one particle, the other view can show that interpretation was erroneous. As already mentioned in reference [15], a multi-view set-up enables an important measurement verification: the sizes of a particle deduced from the out-of-focus images of both views must be the same along the axis that is common to both angles of view. As perspective in flow analysis, we can imagine the seeding of a flow containing irregular particles by metallic microspheres that will play the role of the small particle and will induce the generation of localized holograms (within the out-of-focus image) of nearby irregular rough particles.

**Acknowledgment**. The authors thank the French National Agency for financial support through the CISTIC project of program Investissements d'Avenir LabEx EMC3 (ANR-10-LABX-09-01).